\documentclass[fleqn,10pt]{wlscirep}
\usepackage{epstopdf}
\usepackage{caption}
\usepackage{bm}
\usepackage{subfig}
\usepackage{graphicx}
\usepackage{amsmath,amssymb}

\title{Optical Measurement of In-plane Elastic Waves in Mechanical Metamaterials
Through Digital Image Correlation}

\author[1]{Marshall Schaeffer}
\author[2,*]{Giuseppe Trainiti}
\author[1,2]{Massimo Ruzzene}
\affil[1]{Georgia Institute of Technology, George W. Woodruff School of Mechanical Engineering, Atlanta, 30332, USA}
\affil[2]{Georgia Institute of Technology, Daniel Guggenheim School of Aerospace Engineering, Atlanta, 30332, USA}

\affil[*]{Corresponding author. email: gtrainiti@gatech.edu}

\begin{abstract}
We report on a Digital Image Correlation-based technique for the detection of in-plane elastic waves propagating in structural lattices. The experimental characterization of wave motion in lattice structures is currently of great interest due its relevance to the design of novel mechanical metamaterials with unique/unusual properties such as strongly directional behavior, negative refractive indexes and topologically protected wave motion. Assessment of these functionalities often requires the detection of highly spatially resolved in-plane wavefields, which for reticulated or porous structural assemblies is an open challenge. A Digital Image Correlation approach is implemented that tracks small displacements of the lattice nodes by centering image subsets about the lattice intersections. A high speed camera records the motion of the points by properly interleaving subsequent frames thus artificially enhancing the available sampling rate. This, along with an imaging stitching procedure, enables the capturing of a field of view that is sufficiently large for subsequent processing. The transient response is recorded in the form of the full wavefields, which are processed to unveil features of wave motion in a hexagonal lattice. Time snapshots and frequency contours in the spatial Fourier domain are compared with numerical predictions to illustrate the accuracy of the recorded wavefields and demonstrate the suitability of this technique for the experimental characterization of wave properties. 
\end{abstract}

\begin{document}

\flushbottom
\maketitle
%
%
\thispagestyle{empty}


\section*{Introduction}
Structural lattices belong to the rich and diverse family of mechanical metamaterials and phononic crystals, and have been widely investigated for their superior or unusual mechanical properties~\cite{hussein2014dynamics}. For example, 2D architected materials exhibit buckling-induced effective negative swelling~\cite{Liu2016}, while enhanced resilience and recoverability has been demonstrated for 3D hierarchical designs  versus non-hierarchical ones~\cite{Meza15092015}. The dynamic properties of this class of engineered materials have perhaps received even more attention, mainly given their ability to control the propagation of waves through frequency bandgaps~\cite{phani2006wave}, wave directionality in both periodic and graded topologies~\cite{Casadei2013,Trainiti2016}, negative refraction~\cite{Zhao2016533} as well as topologically protected wave motion~\cite{Kane2013,Mousavi2015,Pal2016top}. Various strategies have been proposed to increase the structural lattices' ability to steer and block energy propagating in the bulk of the material, through reconfiguration of multistable lattices~\cite{Schaeffer2015}, load-induced large deformations~\cite{Pal2016} and active tuning of wave properties through resonant and negative capacitance shunted piezoelectric networks~\cite{Gonella2015,ZhiWei2016}. 

In spite of the intense research focus on lattice mechanics, the experimental documentation of their wave properties has so far been relatively limited. This is due to challenges related to the detection of transient wave phenomena in structural assemblies that are highly porous and that consist of slender elements. Furthermore, characterization often requires the detection of highly resolved wavefields consisting of both transverse and in-plane displacement components. An example of such investigations is presented in~\cite{Celli2014114,Zhu2014} where a 3D Scanning Laser Doppler vibrometer (SLDV) is employed to record multi-component wavefields in an hexagonal lattice. However, SLDVs may have limited applicability in particular as the dimensions of the lattice and of the slender elements composing it is reduced. Thus alternative techniques to achieve similar objectives continue to be of interest. The presentation of one of these techniques is the main objective of the paper. 

This work presents an approach based on the use of high speed cameras along with a Digital Image Correlation (DIC) process, here adapted to allow for the tracking of points belonging to a lattice of known geometry. DIC is a widely used technique mostly for quasi-static analyses, such as mechanical property estimation, strain field measurements and crack specimen studies~\cite{Sutton:2009:ICS:1593508}. Although recently the use of DIC was extended to modal testing~\cite{HAGARA2012180,Wang20111599,Lee201363,Ehrhardt2016}, very little effort has been devoted to date to its application for wave studies. The investigation of wave properties makes the use of DIC challenging, given the void-to-fill ratio of the medium, the high frame rates required and the associated limited field of view.

In the approach presented herein, DIC is employed to detect and track the motion of the lattice intersection points, which are identified through the search of corresponding clusters of pixels. The procedure is currently limited to the detection and monitoring of small displacements that do not cause significant changes in lattice topology and connectivity. The method should be extended to be applicable to investigate the dynamic behavior of structures that undergo large deformations~\cite{Jurjo20101369}, topological changes, and reconfigurations~\cite{Schaeffer2015670}. Challenges associated with the frame rates required for capturing wave motion are here addressed by employing a repetitive excitation that allows for implementing an image interleaving process that artificially enhances the sampling rate and that enables the stitching of subregions of the lattice recorded at subsequent times to enlarge the effective field of view. The obtained wavefield images are then employed for the analysis of the wave motion of the lattice, specifically in terms of the dispersion characteristics of the two in-plane wave modes and of the associated wave velocities.

\section*{Results}
\begin{figure}[h]
\centering
\includegraphics[width=0.7\textwidth]{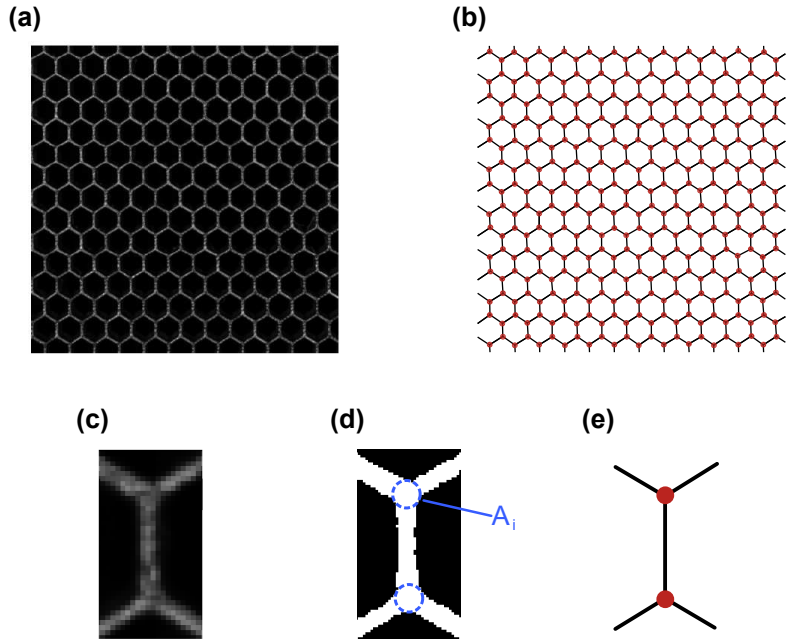}
\caption{Picture of hexagonal lattice (a) and identified lattice geometry consisting of intersection points (red dots) connected by lines (black lines). Schematic of process followed for the identification of intersections and of lattice geometry (c-e).}
\label{Lattice_Schematic}
\end{figure}

The considered DIC approach is applied for the detection of in-plane wave motion of the periodic hexagonal lattice shown in Fig.~\ref{Lattice_Schematic}.a. Motion is evaluated by tracking the position of the lattice intersections, or nodes, in the recorded images. These intersections are identified by setting a brightness threshold to each black and white image of the kind in Fig.~\ref{Lattice_Schematic}.a, which produces a binary representation of material and void as illustrated in Fig.s~\ref{Lattice_Schematic}.c,d. Pixels with brightness above the threshold are denoted as ``material pixels". A tolerance brightness level is applied to connect each material pixel to its neighbors whose brightness may fall just below the threshold, but are still part of the lattice material, which corrects for inaccuracies resulting from the non-uniform illumination of the lattice. The search for lattice intersections proceeds by counting the number of material points within a radius $r=t/\sqrt{3}$ from each material pixel. Here, $t$ denotes the wall thickness of the lattice, so that the radius $r$ corresponds to the theoretical area occupied by an intersection, which is denoted as $A_i$ in Fig.~\ref{Lattice_Schematic}.d. It is expected that the material points with the largest number of neighbors within radius $r$ correspond to the location of the intersection. The process is guided by the approximate estimate of the ratio of the area occupied by an intersection relative to the area of a unit cell. Based on the schematic of Fig.~\ref{Lattice_Schematic}.d, the area occupied by three beams connected at one node is $A \approx 3/2 l t$, where $l$ denotes the length of a ligament and the distance between neighboring nodes, while the area occupied by an intersection is $A_i \approx \pi t^2 / 3$. The considered aspect ratio $l/t \approx 3.725$ gives $A_i / (A+A_i) \approx 15.8\%$. This ratio is employed for an initial estimate of the number of pixels, out of those identified from the thresholding, that are expected to define an intersection. This target number is lowered by an 80\% factor to account and correct for variability in the thresholding process related to imperfections in manufacturing, limitations in image resolution and slight inhomogeneities in illumination. This value is chosen empirically based on comparisons of identified geometry and original images, and is found to provide accurate detection of all nodes. Finally, the location of the intersection is defined by the centroid of the material points within each cluster of radius $r$ . Once the intersections are evaluated, their connectivity is determined by looking for intersections which are approximately $l$ apart. This results in the geometrical description of the lattice as an assembly of points connected by lines as shown in Fig.~\ref{Lattice_Schematic}.e at the local level, and in Fig.~\ref{Lattice_Schematic}.b for the lattice assembly. This simple procedure is limited to the specific lattice topology considered, and will require extension to handle a variety of topologies with complex connectivities and a variety of intersections characterized by different coordination numbers.

\begin{figure}[h]
\centering
\includegraphics[width=0.8\textwidth]{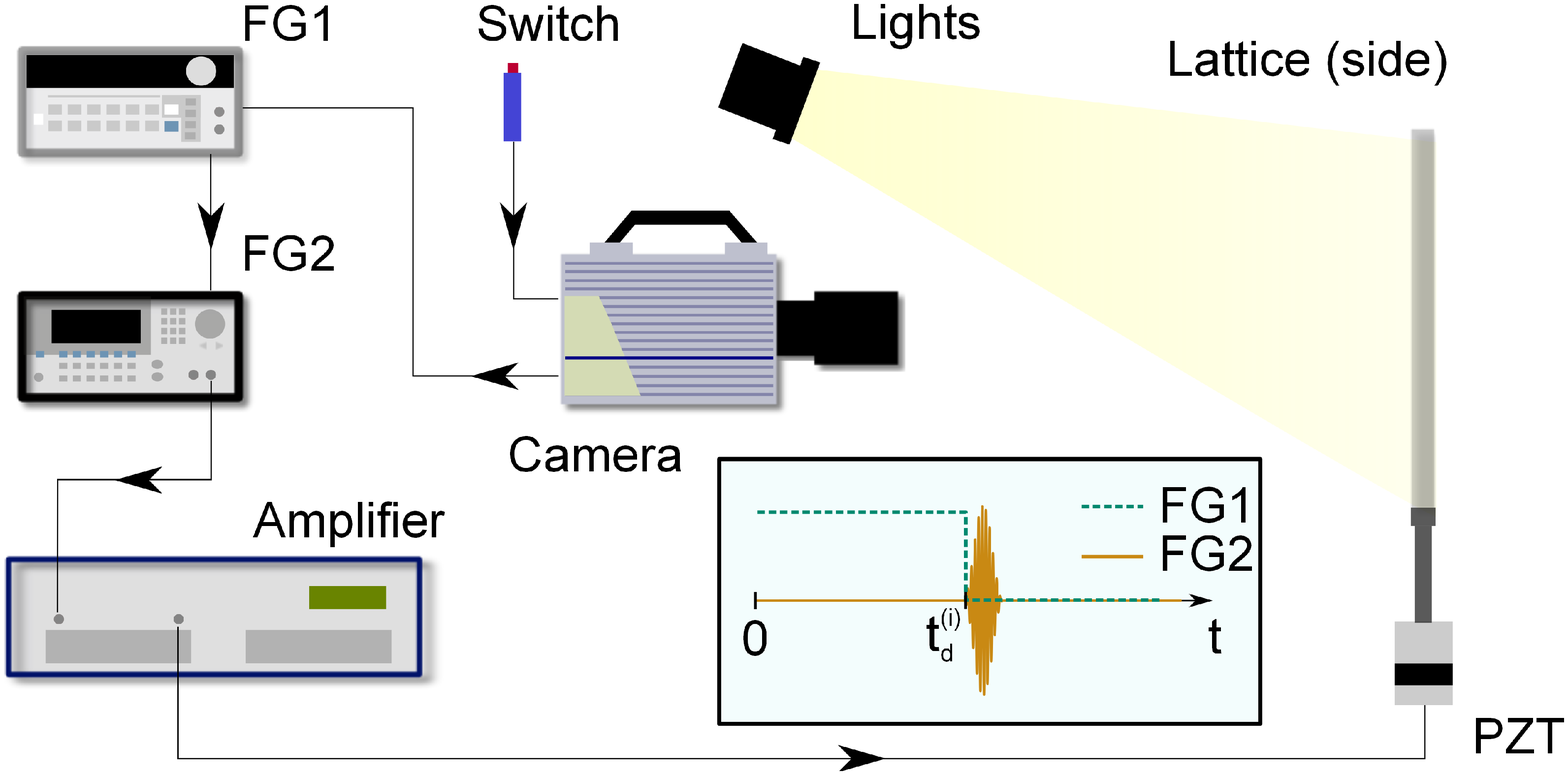}
\caption{Schematic of experimental set-up.}
\label{Experimental Set-up}
\end{figure}

\begin{figure}[h]
\centering
\includegraphics[width=0.75\textwidth]{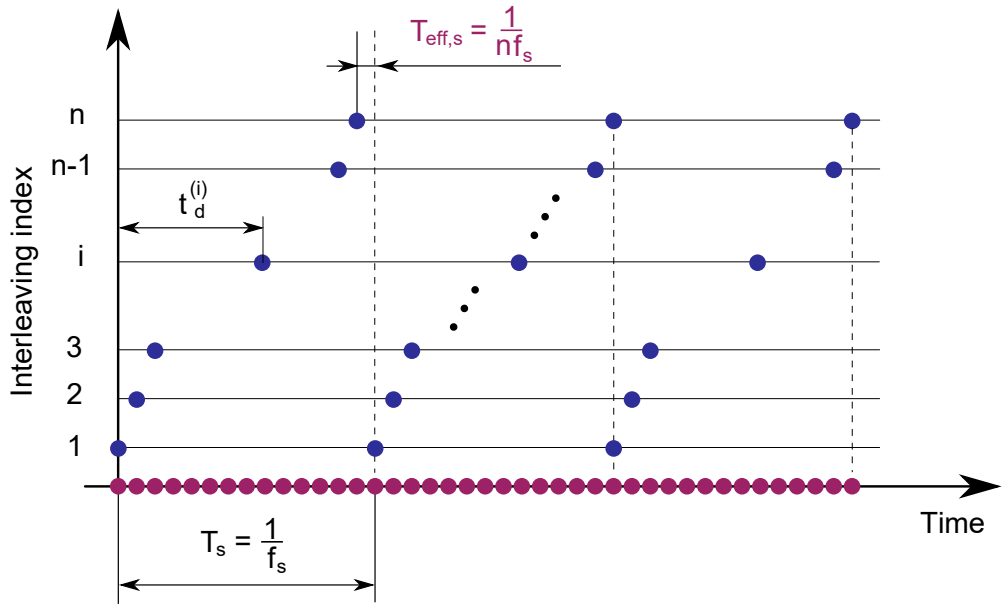}
\caption{Schematic of interleaving process for enhancement of effective sampling rate.}
\label{Interleaving}
\end{figure}

\begin{figure}
\centering
\includegraphics[width=0.75\textwidth]{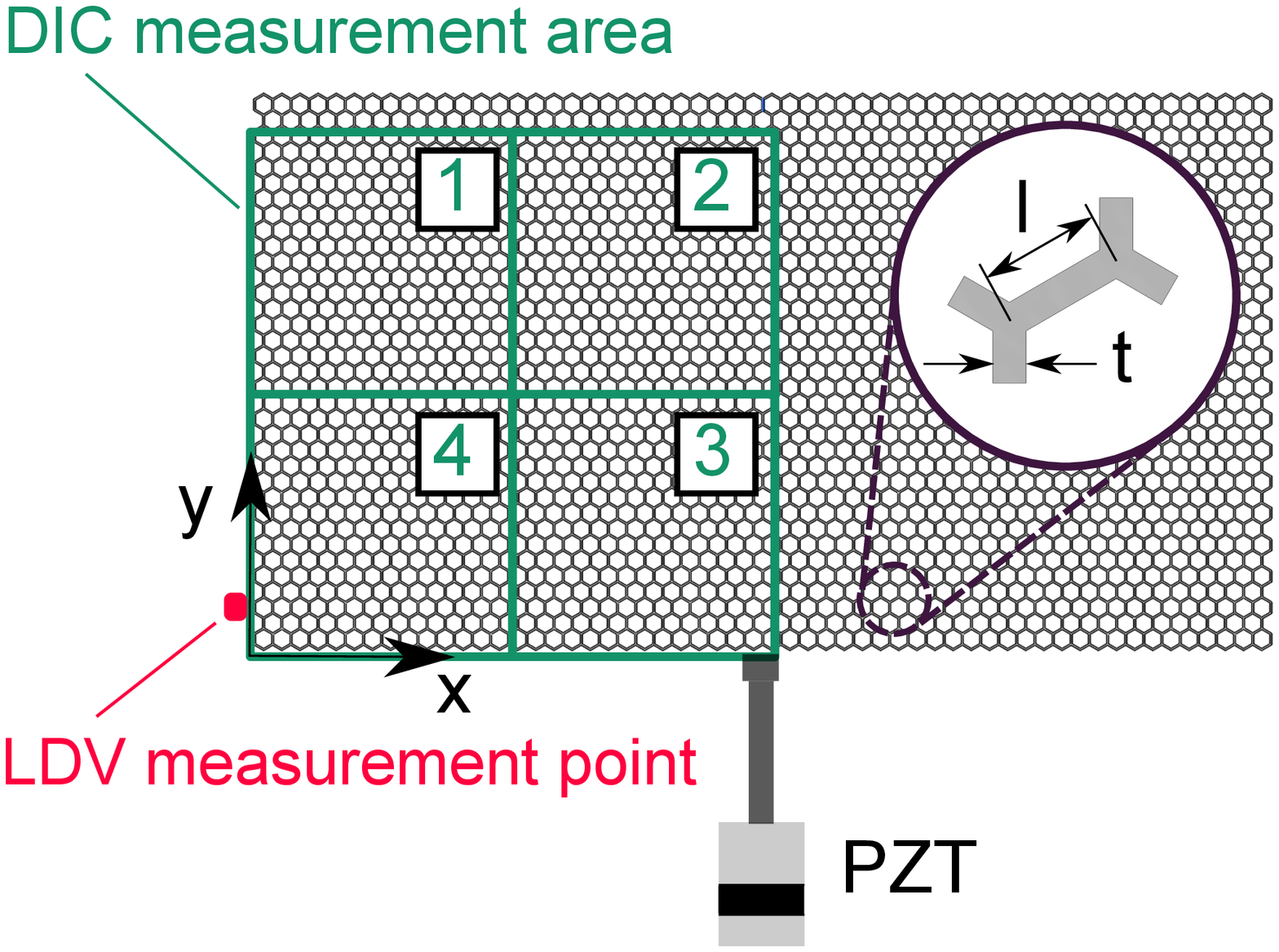}
\caption{Schematic of monitored portion of surface area of the lattice, and subdivision into 4 tiles. The figure also illustrates the point of excitation as well as the location of the LDV measurements.}
\label{Tiles}
\end{figure}

The intersection identification process is applied to the first image recorded during the acquisition time, so that displacements can be evaluated through DIC~\cite{eberl2010free,JonesSilbersteinWhiteEtAl2014}. The DIC process in this work employs the open-source code by Eberl et al.~\cite{Jones2013}, here adapted to track the intersection points. DIC applied to 19x19 pixel image subsets that are located at each lattice intersection, which provides a measure of the two in-plane displacement components relative to the initial reference lattice position. The recorded motion corresponds to the transient wave induced in the lattice by a piezoelectric (PZT) stacked disks assembly designed to resonate at a specified frequency. The motion of the structure is recorded by a high-speed camera which is triggered by the excitation signal. A schematic of the experimental set-up is depicted in Fig.~\ref{Experimental Set-up}, while details of the equipment and excitation parameters are provided in the `Methods' section. The excitation system is selected to provide repeatable forcing and, as a result, repeatable wave motion in the lattice. This has two main advantages. First, hardware limitations of the camera impose a trade-off between frame rate and image size, \emph{i.e.} the number of pixels captured in each frame.  To circumvent this, images from $n$ recordings at a frame rate of $f_s$ are interleaved to obtain an effective higher frame rate $f_{s,eff} = nf_s$. This is implemented by controlling the time delay between the beginning of a recording and the start of an excitation. The time delay corresponding to the $i$-th recording is given by:  
\begin{equation}
t_d^{(i)} = \frac{1}{f_s}\frac{i-1}{n} 
\end{equation}
where $i\in{1,...,n}$, with $n$ denoting is the number of the recordings. In this work, we used $n = 14$ delayed recordings to realize an effective $f_{s,eff} = 14 \times f_s$, where $f_s=5$~kHz is the frame rate provided by the video camera utilized. A schematic of the interleaving process is presented in Fig.~\ref{Interleaving}. A second advantage is that it provides a wide effective field of view that allows capturing a sufficiently large portion of the surface area of the lattice. We elected to capture the motion only of half of the lattice, invoking symmetry of geometry and loading configuration. The monitored half surface is further divided into 4 tiles as illustrated in Fig.~\ref{Tiles}. Recordings are first conducted on each of the tiles. A composite video is then obtained from the combination of the 4 tiles which is obtained by aligning the intersection locations in planned overlapping regions.  The composite images are obtained by first rotating each tile's coordinate system so that the edges of the overlapping area align with the global coordinate axes.  Then, each tile is translated so that the centroid of its overlap region matches that of its pairing neighbor.  

\begin{figure}[h]
\centering
\includegraphics[width=0.65\textwidth]{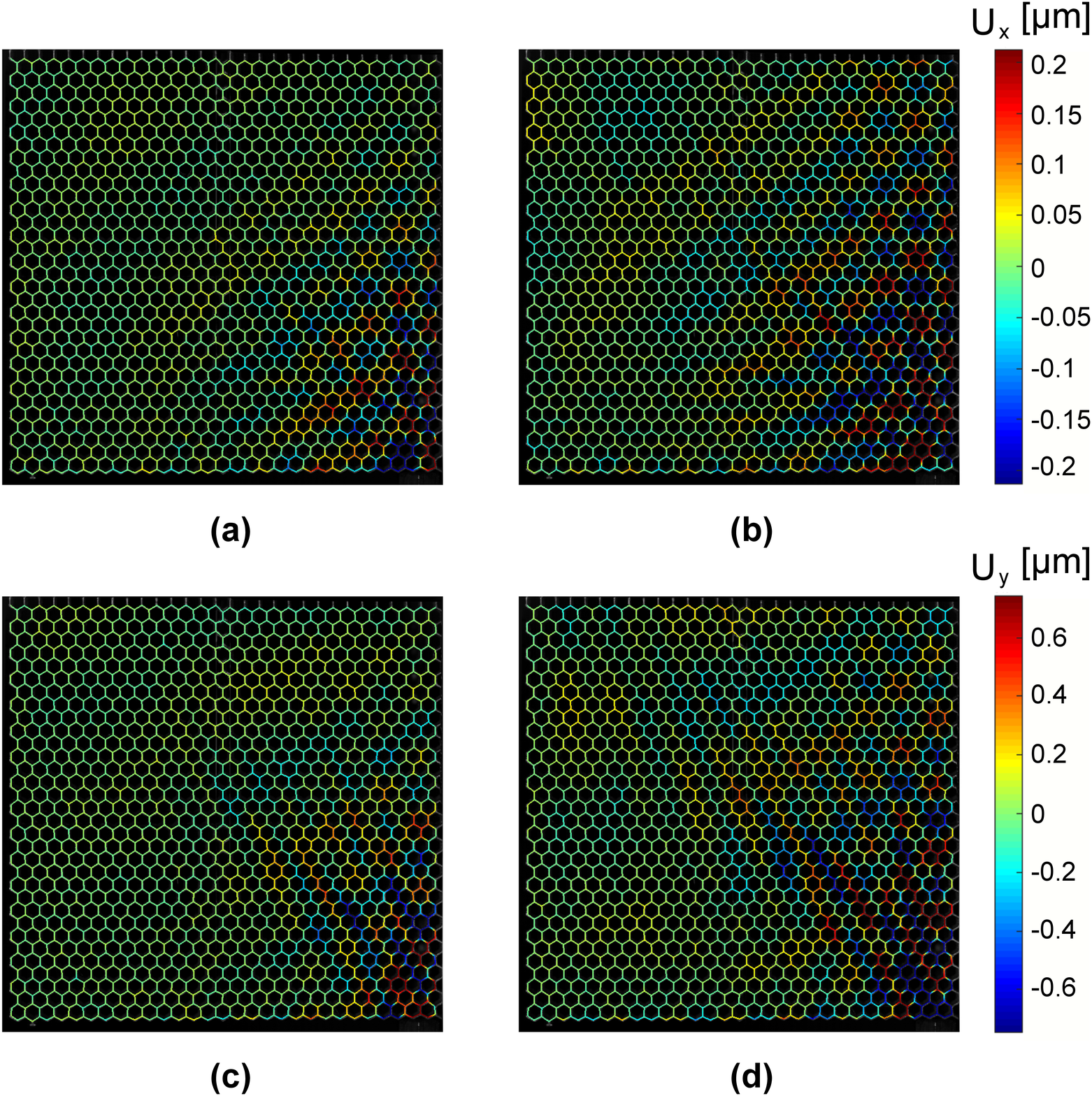}
\caption{Time snapshots of the recorded wave motion in the lattice resulting from the DIC process presented: $t=3.78$ ms (a,c), $t=3.99$ ms (b,d) (Horizontal $x$ component (a,b), Vertical $y$ component (c,d)).}
\label{Time snapshots}
\end{figure}

The recorded motion of the lattice can be represented in the form of the time snapshots in Fig.s~\ref{Time snapshots}, where the color code is associated with the resultant of the in-plane displacement components .

\begin{figure}[h]
\centering
\includegraphics[width=0.4\textwidth]{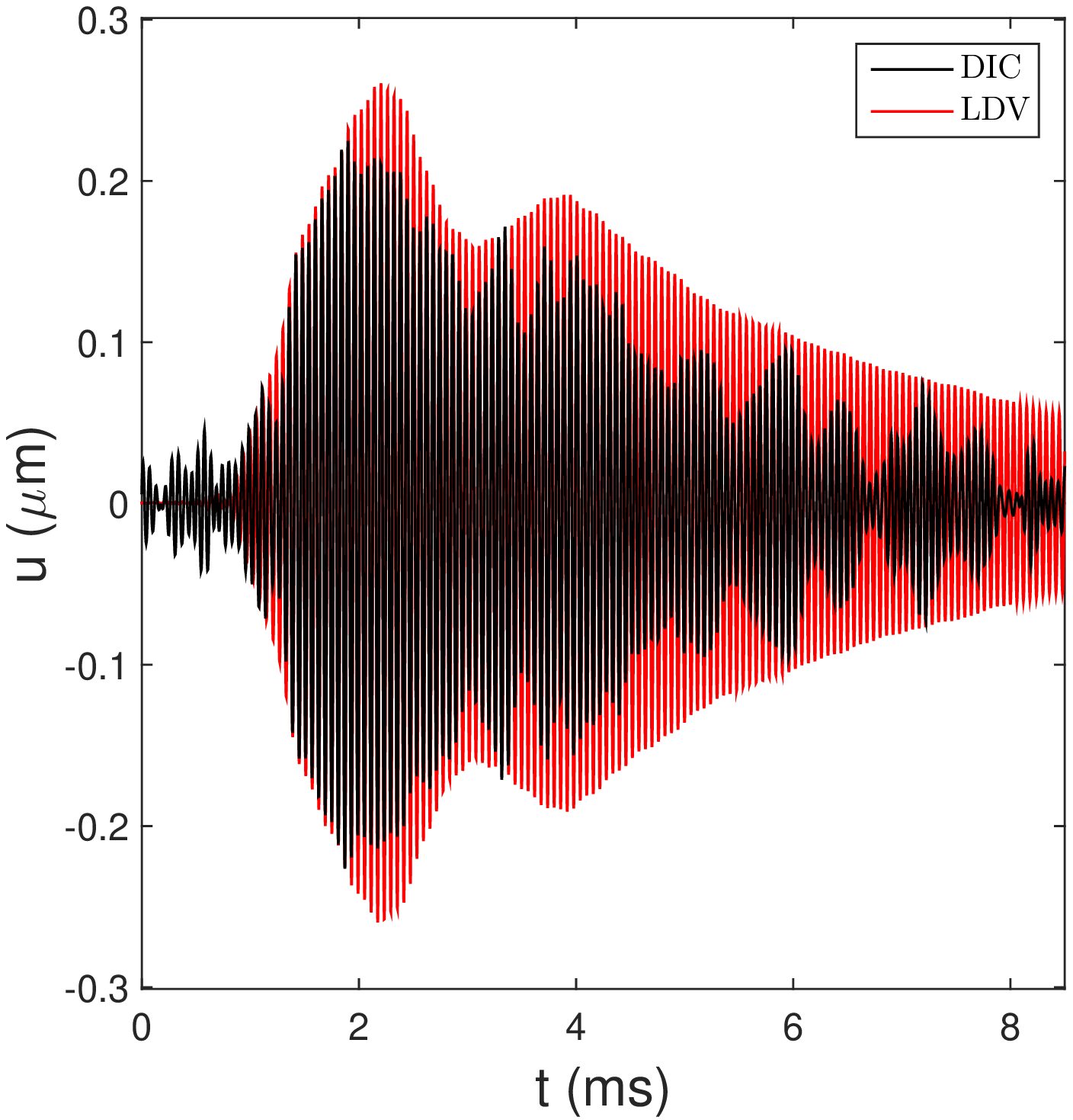}
\caption{Comparison of DIC time trace with LDV measurements recorded at the location shown in Fig.~\ref{Tiles}.}
\label{LDVvsDIC}
\end{figure}

In addition, the motion of individual points can be extracted from the recorded images to obtain individual time-traces that can be compared with point measurements obtained for example from a single point LDV. An example of such a comparison is shown in Fig.~\ref{LDVvsDIC} to illustrate how the method proposed herein is capable of providing information that is comparable with that recorded by the LDV. While the results from the DIC (black solid line) appear more noisy, which is evident from the recorded signal prior to the arrival of the wave, which occurs at $t\approx 0.5$ ms, they compare well in general trends in terms of amplitude, and of rising and decaying trends. Of note is the fact the the LDV's data were obtained from numerical integration of the LDV velocity outputs. In addition, imperfect alignment of the LDV beam relative to the lattice may lead to contribution of both horizontal and vertical displacement component, which could be considered as reasons for the relative mismatch between the data. The comparison of the two time traces indicates how the DIC technique can be employed for the description of wave motion through measurements that contain sufficient information in space and time for subsequent characterization of the wave mechanics of the lattice as discussed in the upcoming ``Discussion" section.

\section*{Discussion}
Analysis of the wave motion of the lattice as recorded through the process described in the previous section reveals some of the interesting properties that characterize the hexagonal topology here considered. The results also suggest how this measurement approach may be applied to investigate the properties of a variety of lattice topologies. Examples of such investigation enabled by the in-plane detection of wave motion are described in the following paragraphs.

The time snapshots of Fig.s~\ref{Time snapshots} illustrate how the response of the lattice at the considered excitation frequency ($f_{e}=16.74$ kHz - see ``Methods" Section) is directional in space, i.e. wave motion occurs preferentially along the vertical ($y$) direction. This is the result of the anisotropy of this lattice which is documented for example in~\cite{Gonella2008125}. Such directionality can be predicted numerically by performing the dispersion analysis of the lattice as described for example in~\cite{Trainiti2016}. The experimental determination of the directional properties however is complicated by the fact that the recorded response is a superposition of the in-plane wave modes corresponding to longitudinal (P-mode) and transverse (S-mode) polarizations.

\begin{figure}[ht]
\centering
\subfloat[]{\includegraphics[width=0.4\textwidth]{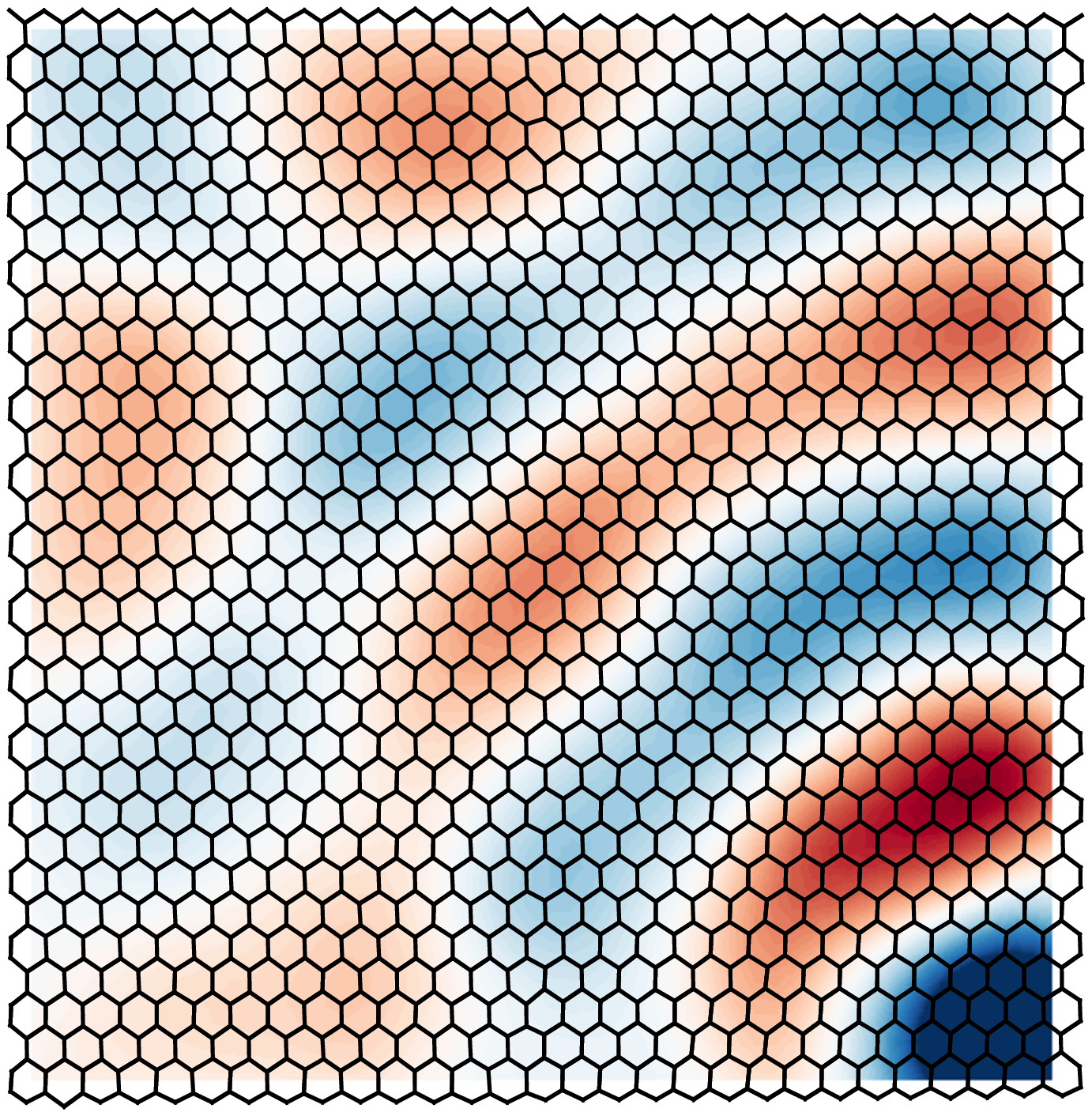}}
\subfloat[]{\includegraphics[width=0.4\textwidth]{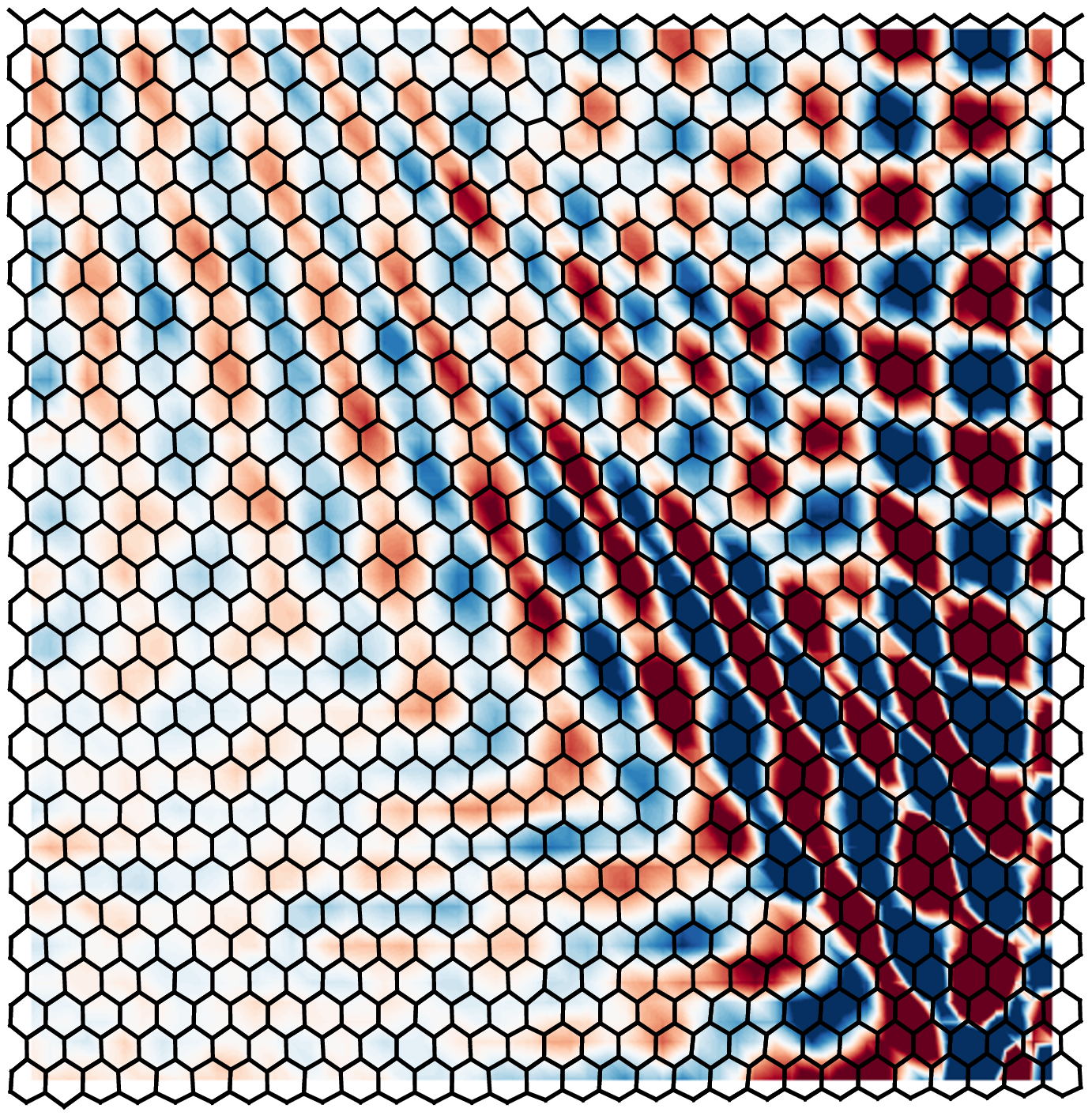}}
\caption{Snapshot of divergence  $\mathcal{P}(x,y,t)$ (a) and curl $\mathcal{S}(x,y,t)$ (b) of the measured displacement field at $t=3.78$ ms and $t=4.14$ ms, respectively.}
\label{PS_mode_time}
\end{figure}

\begin{figure}
\centering
\subfloat[]{	\includegraphics[width=0.4\textwidth]{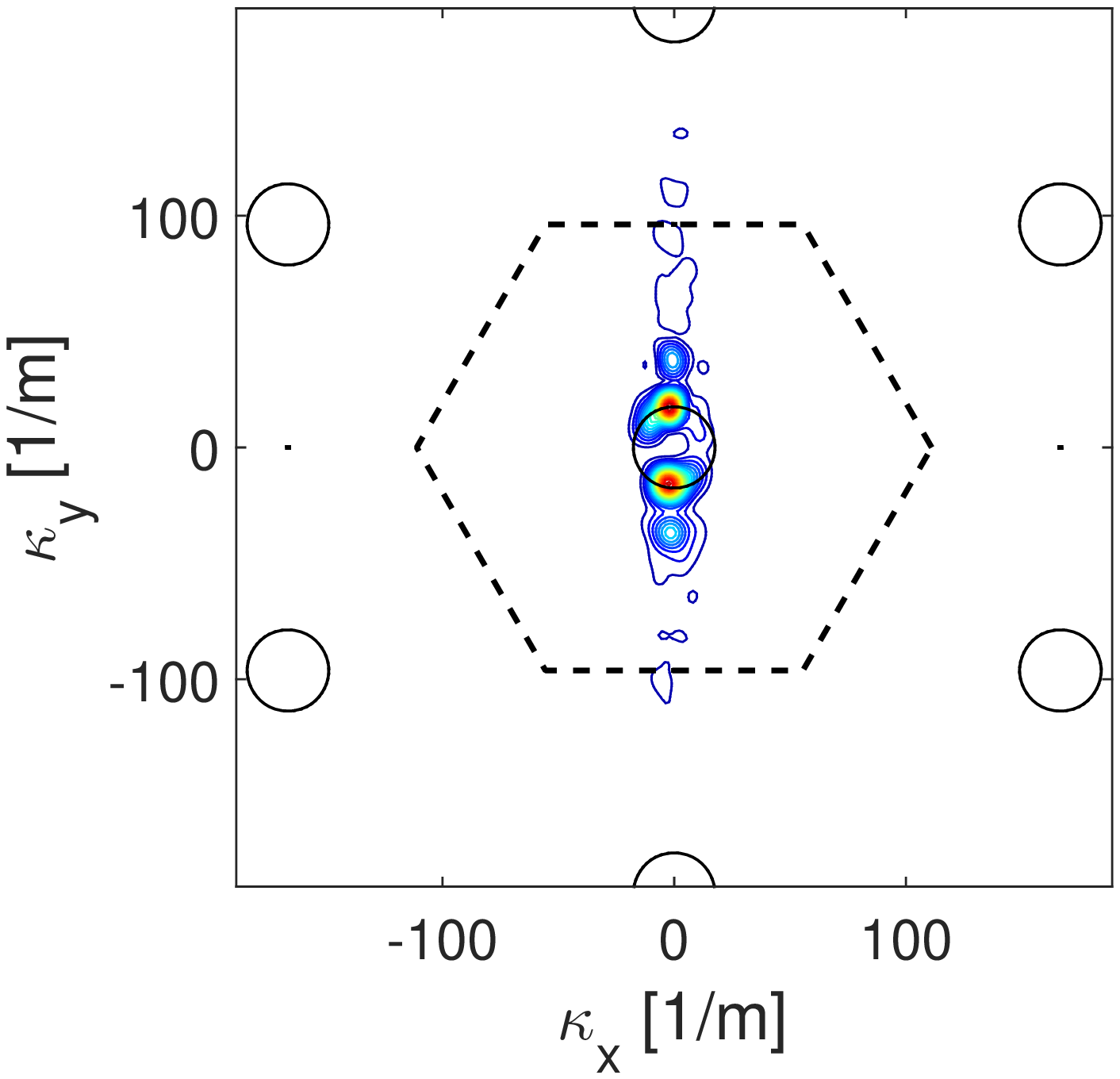}}
\subfloat[]{	\includegraphics[width=0.4\textwidth]{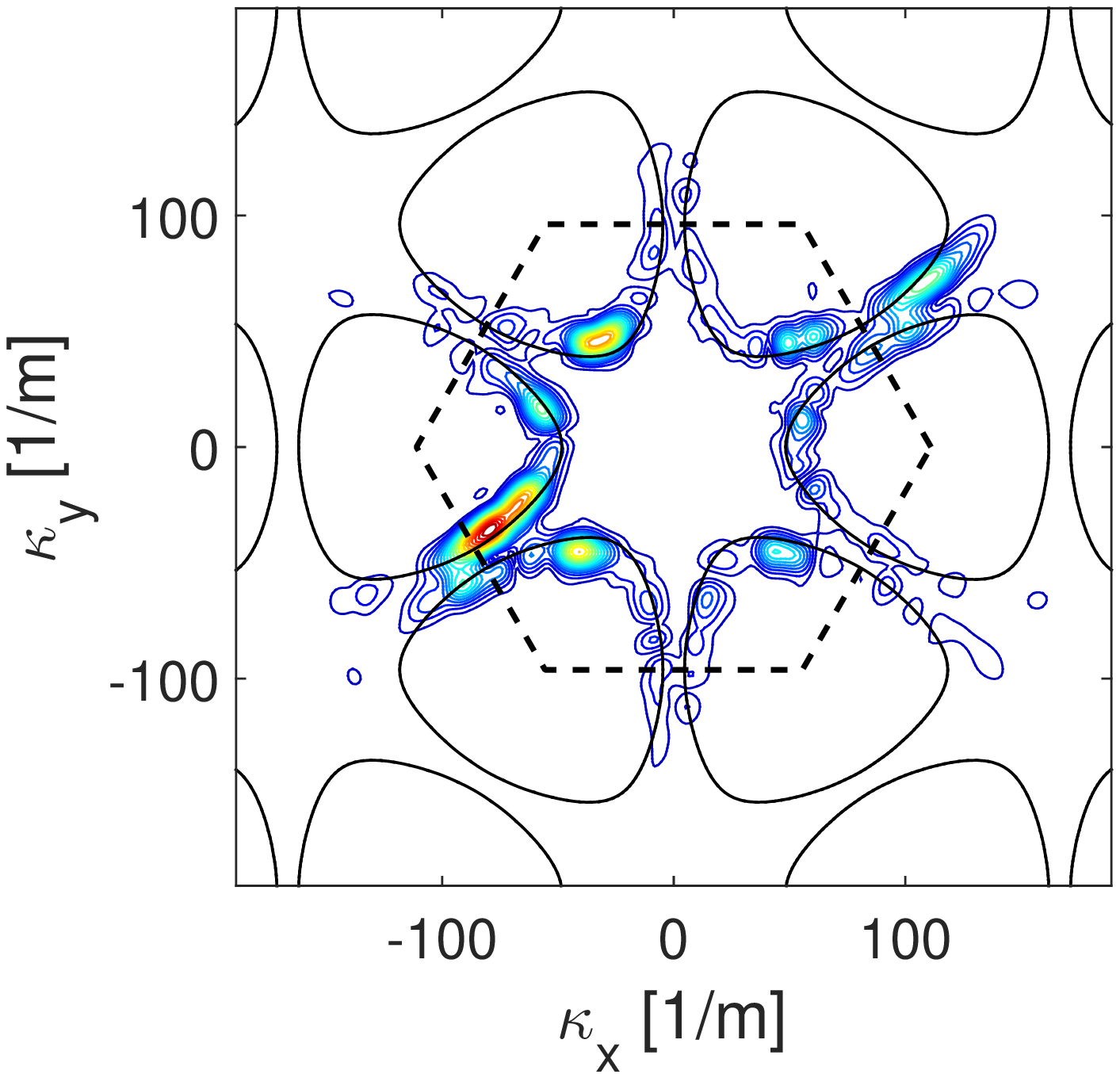}}
\caption{Wavenumber domain representation $|\mathcal{\hat{P}}(k_x,k_y,\omega)|$ (a), and $|\mathcal{\hat{S}}(k_x,k_y,\omega)|$ (b) at the excitation frequency and comparison with theoretical iso-frequency dispersion contours (solid black lines). The dashed black line outlines the First Brillouin Zone of the reciprocal space for the lattice~\cite{hussein2014dynamics}\cite{kittel2004introduction}.}
 \label{PS_mode_freq}
 \end{figure}

The availability of the in-plane displacement vector $\bm{u} = u \bm{i} + v \bm {j}$, with $\bm{i}, \bm {j}$ denoting the unit vectors aligned with the horizontal ($x$) and vertical ($y$) axes respectively, however affords the possibility to effectively separate the two modal contributions. This is achieved by employing Helmholtz decomposition to express the displacement $\bm{u}$ as a function of the scalar P-wave potential $\phi$ and the vector S-wave potential $\bm{\psi}$~\cite{Casadei2013}:
\begin{equation}\label{eq: helmholtz decomposition}
\bm{u} = \nabla\phi + \nabla\times\bm{\psi}
\end{equation}
where the divergence $\nabla\cdot\bm{u} = \nabla^2\phi$ and the nonzero component of the curl $\nabla\times\bm{u} = -\nabla^2\psi$ of the wavefield, in this case directed along the out-of-plane direction $z$, separate P and S-wave contributions, respectively. The spatial derivatives required for the computation in eq.~\ref{eq: helmholtz decomposition} are computed by first interpolating the displacements of the hexagonal grid onto a rectangular grid. This is based on a natural neighbor interpolation routine available in the Matlab environment. Next, a central difference scheme is employed for the computation of derivatives. The time snapshots of $\mathcal{P}(x,y,t)=\nabla\cdot\bm{u} $ and of $\mathcal{S}(x,y,t)=\nabla\times\bm{u}|_z$ shown in Fig.~\ref{PS_mode_time} illustrate the differences in speed, wavelength and directionality between the two wave modes. In both figures, the color map is based on normalized data, with red and blue respectively denoting maximum positive and negative values. Most notably, and as expected, the P-mode has a significantly larger wavelength and appears more isotropic than the S-mode, which is anisotropic with preferential propagation along the positive $y$ direction and towards the top left corner of the  structure. The maps of the separated modes also show that the S-mode dominates the overall response shown in Fig.~\ref{Time snapshots}. Directionality and wave speeds can be characterized and quantified by representing the two modal components $\mathcal{P}(x,y,t), \mathcal{S}(x,y,t)$  in the Fourier domain, i.e. $\mathcal{\hat{P}}(k_x,k_y,\omega), \mathcal{\hat{S}}(k_x,k_y,\omega)$ at a specific frequency. The contour plots of Figs.~\ref{PS_mode_freq} correspond to the magnitudes $|\mathcal{\hat{P}}(k_x,k_y,\omega)|, |\mathcal{\hat{S}}(k_x,k_y,\omega)|$ at the excitation frequency $f_e=16.74$ kHz and provide a distribution map of the energy content of the recorded response in the wavenumber domain. The maps also superimpose the iso-frequency contours of the theoretical dispersion surfaces for the lattice, which are represented by the solid black lines in both Fig.s~\ref{PS_mode_freq}. The theoretical dispersion relations are evaluated by applying established methods based on the Finite Element (FE) discretization of a unit cell, and the application of Bloch periodic conditions~\cite{Trainiti2016}. Both maps in Fig.s~\ref{PS_mode_freq} show a good match between theoretical predictions and measured data, which demonstrates that the detected motion is descriptive of the wave properties of the lattice, and that the representation of the measured wavefields in the wavenumber space can be an effective tool for the estimation of the dispersion properties, and therefore the mechanical properties of a lattice under investigation. Of note is the fact that the iso-frequency contour has the form of a circle for the P-mode (Fig.~\ref{PS_mode_freq}.a), which matches the theoretical predictions well, while the S-mode reflects more predominantly the hexagonal geometry of the lattice and its six-fold symmetry (Fig.~\ref{PS_mode_freq}.b).
\begin{figure}
\centering
\includegraphics[width=0.4\textwidth]{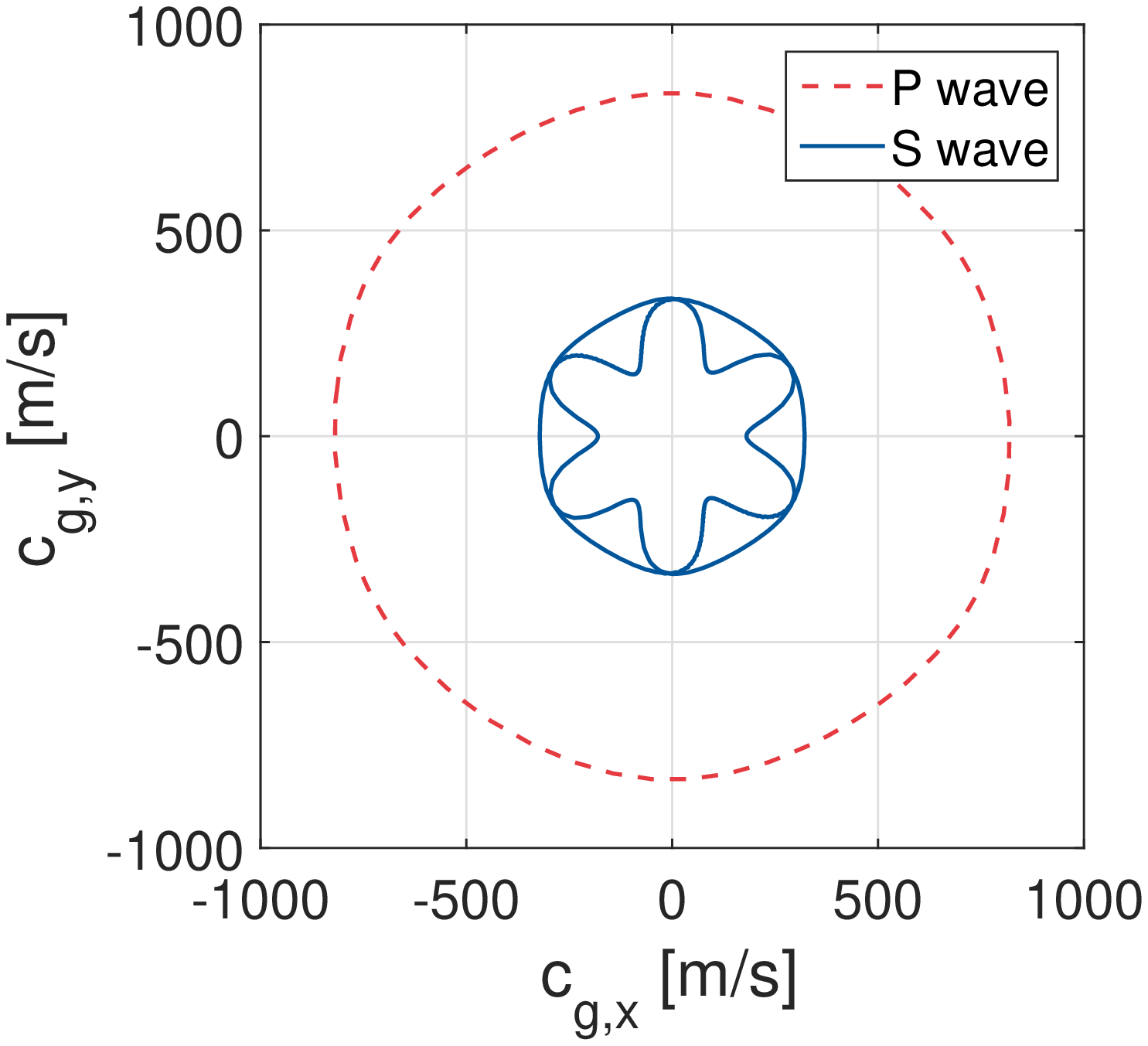}
\caption{Direction variations of group velocity at the excitation frequency $f_{e}=16.74$ kHz: S-mode: solid blue line, P-mode: dashed red line.}
\label{Group_velocity}
\end{figure}
The theoretical predictions of dispersion can be further elaborated to obtain the group velocities for the two modes at designated frequencies. Theoretical group velocities for the two modes are shown in Fig.~\ref{Group_velocity} illustrate their different anisotropy, and further highlight the directional properties of the S-mode in particular. The group velocity directional variations provide an estimate of the shape of the wave front corresponding to the propagation in the lattice of a wave packet. Thus, at a particular time, one may superimpose the predicted wavefront to the one actually measured to once again illustrate the capabilities of the method presented here along with the theoretical estimation of dispersion to investigate wave propagation characteristics in terms of both magnitude of group velocity and directional dependence. Specific to the results in Fig.~\ref{Group_velocity}, isotropy of the P-mode at $f_{e}=16.74$ kHz is confirmed by the corresponding group velocity plot, which is almost a perfect circle, signaling that the energy carried by the P-mode spreads in all the direction, with about the same average speed approximately equal to 820 m/s. The ``star'' pattern shown in Fig.~\ref{PS_mode_freq}.b instead captures the anisotropic propagation of the S-waves, which is in good agreement with theoretical predictions showing directions along which the energy is focused by the S-wave with 6-fold symmetry, with faster components of the wave traveling at about 320 m/s.

\section*{Methods}
\subsubsection*{Lattice Fabrication}
The hexagonal honeycomb lattice is manufactured out of polylactic acid (PLA) using fused deposition manufacturing (FDM). The lattice is characterized by 56x36 unit cells of dimensions $l=3$ mm, $t=0.8053$ mm  (Fig.~\ref{Lattice_Schematic}.a), and out-of-plane width $w=10$ mm. The number of cells is maximized in relation to the overall area than can be fabricated by available machines. The beams comprising the hexagons have slenderness ratio $l/t\approx3.73$ which leads to a fairly small material to void ratio.  A densely filled region on the lattice boundary is included to facilitate fixturing and approximate a rigid boundary.  A reflective mirror spray paint is applied to the lattice, enhancing its ability to reflect the bright lights required for short camera exposures and protects the structure from overheating.  Before painting, the structure is sanded with 240 grit sandpaper to give more defined features for DIC to utilize. 

\subsubsection*{Experimental setup}
The experimental setup is shown in Fig.~\ref{Experimental Set-up}.  The excitation to the lattice is provided by a resonant piezoelectric (PZT) stack assembly (APC 90-4060) attached to the lattice in the location shown. The PZT resonance frequency is the considered excitation frequency $f_{e}=16.74$ kHz. The excitation is applied in the form of a sinusoidal burst, with modulation being provided by an $11$-cycle Tuckey window. The motion of the structure is recorded by a high-speed camera (Photron Fastcam SA1.1) which is set to record images at a rate of $f_s = 5,000$ frames per seconds. The interleaving process described above considers $n=14$ properly delayed takes, so that the effective sampling frequency is $f_{s,eff} = n f_s = 70k$Hz. Each recording is repeated $16$ times and averaged to minimize noise.  Recording is started by a manual switch connected to the camera. Once the recording started, the camera triggers a first function generator (FG1 - Agilent 33120A), which produces a square wave. The falling edge of the square wave then triggers the excitation signal produced by second function generator (FG2 - Agilent 33220A). Imposing the frequency of the first signal generator allows us to enforce the time delay needed to achieve the effective sampling rate $f_{s,eff}$. The excitation signal is then fed to the PZT exciter through an amplified (E\&I 1040L).  The motion of one point of the lattice is also monitored through a Laser Doppler Vibrometer (LDV) (Polytec PDV 100). The location of the monitored point is shown also in Fig.~\ref{Experimental Set-up}. The LDV measurements are employed to verify the repeatability of the excitation and motion during various takes and excitation cycles, as well as to compare the time histories recorded at the monitored location with those extracted from the DIC procedure. During data collection, uniform illumination of the lattice is provided by two lamps. These lamps are observed to have a small but detectable flicker in brightness at 120~Hz which is filtered out by a high-pass filter applied before the interleaving. In addition, mean subtraction in time is applied to remove any low frequency motion that may contribute to noise. A Hamming window is then applied in the time domain to minimize spectral leakage. 

\subsubsection*{FE Modeling and Bloch Analysis}
Bloch analysis is implemented numerically by using the FE commercial code ABAQUS/Standard~\cite{aaberg1997usage}. The unit cell, represented in Fig.~\ref{Tiles}, is discretized by using three-dimensional hexahedral C3D8R ABAQUS elements. The Bloch conditions, depending on the  wavenumber $\bm{\kappa}$, are enforced on the nodal displacement of the boundaries of two identical unit cells, defined as real and imaginary unit cells. An eigenvalue problem is solved for the frequency $\omega$ for each value of $\bm{\kappa}$ used to discretize the Irreducible Brillouin Zone in order to obtain the dispersion relation $\omega=\omega(\bm{\kappa})$ for the structure. The material is modeled as linear elastic with Young's modulus $E=2.46$ GPa, obtained by previously gained empirical knowledge, mass density $\rho=1250$ $kg/m^3$ and Poisson's ratio $\nu=0.36$.
\bibliography{references}

\section*{Acknowledgements}
This work is equally supported by grants from the Army Research Office (Grant Number: W911NF1210460), and from the Air Force Office of Scientific Research (Grant Number: FA9550-13-1-0122).

\section*{Author contributions statement}
M.S. conducted the experiments and the post-processing of the data, G.T. contributed to data post-processing and conducting the numerical evaluations of the dispersion properties, and M.R. oversaw the activities and contributed to the drafting of the manuscript.

\section*{Additional information}
The are no competing financial interests from any of the authors of the paper.

\end{document}